\documentclass[iop]{emulateapj}
\usepackage{epsfig}
\usepackage{apjfonts}
\usepackage{aas_macros}
\usepackage{amsmath}
\usepackage{graphicx}
\usepackage{color}
\usepackage{epstopdf}

\begin{document}

\title{Relation between gravitational mass and baryonic mass for non-rotating and rapidly rotating neutron stars}
\author{He Gao$^{1,*}$, Shun-Ke Ai$^{1}$, Zhou-Jian Cao$^{1}$, Bing Zhang$^{2}$, Zhen-Yu Zhu$^{3,4}$,  Ang Li$^{3}$, Nai-Bo Zhang$^{5}$ and Andreas Bauswein$^{6}$}
\affiliation{
$^1$Department of Astronomy, Beijing Normal University, Beijing 100875, China; gaohe@bnu.edu.cn
\\$^2$Department of Physics and Astronomy, University of Nevada Las Vegas, NV 89154, USA;\\
  $^3$Department of Astronomy, Xiamen University, Xiamen, Fujian 361005, China;\\
  $^4$Institute for Theoretical Physics, Max-von-Laue-Strasse 1, 60438 Frankfurt, Germany;\\
  $^5$Shandong Provincial Key Laboratory of Optical Astronomy and Solar-Terrestrial Environment, School of Space Science and Physics, Institute of Space Sciences, Shandong University, Weihai, 264209, China;\\
  $^6$GSI Helmholtzzentrum f\"ur Schwerionenforschung, Planckstra{\ss}e 1, 64291 Darmstadt, Germany.
  }

\begin{abstract}
With a selected sample of neutron star (NS) equation-of-states (EOSs) that are consistent with the current observations and have a range of maximum masses, we investigate the relations between NS gravitational mass $M_g$ and baryonic mass $M_b$, and the relations between the maximum NS mass supported through uniform rotation ($M_{\rm max}$) and that of nonrotating NSs ($M_{\rm TOV}$). We find that if one intends to apply an EOS-independent quadratic, universal transformation formula ($M_b=M_g+A\times M_{g}^2$) to all EOSs, the best fit $A$ value is 0.080 for non-rotating NSs only and 0.073 when different spin periods are considered. The residual error of the transformation is as large as $\sim0.1M_{\odot}$. For different EOSs, we find that the parameter $A$ for non-rotating NSs is proportional to $R_{1.4}^{-1}$ (where $R_{1.4}$ is NS radius for 1.4$M_\odot$ in unit of km). For a particular EOS, if one adopts the best-fit parameters for different spin periods, the residual error of the transformation is smaller, which is of the order of 0.01$M_\odot$ for the quadratic form and less than  0.01$M_\odot$ for the cubic form ($M_b=M_g+A_1\times M_{g}^2+A_2\times M_{g}^3$). We also find a very tight and general correlation between the normalized mass gain due to spin $\Delta m\equiv(M_{\rm max}-M_{\rm TOV})/M_{\rm TOV}$ and the spin period normalized to the Keplerian period ${\cal P}$, i.e. ${\rm log_{10}}\Delta m = (-2.74\pm0.05){\rm log_{10}}{\cal P}+{\rm log_{10}}(0.20\pm 0.01)$, which is independent of EOS models. Applications of our results to GW170817 is discussed.
\end{abstract}


\section {Introduction}

The structure of neutron stars (NSs) depends on the poorly understood physical properties of matter under extreme conditions, especially the equation-of-state (EOS) of matter at the nuclear density \citep{lattimer12}. In NS problems, two masses, i.e. the baryonic mass $M_b$ and the gravitational mass $M_g$, are usually discussed. The former ($M_b$) is theoretically relevant, since it is directly connected to the mass of the iron core of the progenitor massive star. In the problems of NS-NS mergers, it is the baryonic mass that is conserved. The latter ($M_g$), on the other hand, is directly measured from observations, and is smaller than $M_b$ due to the subtraction of the binding energy. Studying the general relationship between $M_g$ and $M_b$ becomes ever more important, as the accuracy of NS mass measurements become sufficiently precise to 1) set interesting lower limits on the NS maximum mass $M_{\rm TOV}$ of nonrotating stellar models, and thus ruling out some soft EOS models \citep{lattimer07,lattimer12}; 2) to study the NS initial mass function \citep{timmes96,fryer12,pejcha15}; 3) to study the neutrino emission from core collapse supernovae \citep{lattimer89,burrows92,gava09,camelio17}, and so on. 

In the literature, the relationships between $M_g$ and $M_b$ or between binding energy $M_b-M_g$ and $M_g$ have been investigated by many authors, most of which have focused on nonrotating non-magnetized NSs at zero temperature \cite[e.g.][]{lattimer89,lattimer01,timmes96,coughlin17}, with some studying rotating NSs \cite[e.g.][]{bozzola18}. 

The effect of NS spin is important in some physical problems. The most relevant problem is binary neutron star (BNS) mergers. For example, one may place important constraints on the NS maximum mass \citep{lasky14,ravi14,lawrence15,fryer15,lv15,gao16,li16,margalit17,rezzolla18,ruiz18} if the gravitational mass of the BNS merger remnant, its rotational properties and its life time until black-hole formation can be inferred from observations. These ideas typically involve the conversion from gravitational mass to baryonic mass or vice versa for either non-rotating NSs or rotating stellar objects. Specifically speaking, it is generally assumed that NS-NS mergers conserve the baryonic mass, with $M_{\rm ej}\lesssim 10^{-1}M_{\odot}$ of materials ejected during the merger \cite[e.g.][]{rosswog13,bauswein13,hotokezaka13,fernandez15,song18}. On the other hand, GW detection can only provide the total gravitational mass of the system at infinite binary separation, $M_{\rm g,tot}$. In order to calculate the mass of the merger remnant, one needs to first convert $M_{\rm g,tot}$ to total baryonic mass $M_{\rm b,tot}$, then convert $M_{\rm b,tot}-M_{\rm ej}$ back to remnant gravitational mass $M_{\rm g,rem}$. According to the observed galactic NS binary population, before merger, the relatively low spin period is expected for both NSs in the binary \cite[][for details]{abbott17}. Therefore, the relation between $M_g$ and $M_b$ for non-rotating NSs should be enough when converting $M_{\rm g,tot}$ into $M_{\rm b,tot}$.  However, since the newborn central remnant must be rapidly spinning, if the remnant is a uniformly rotating NS, the conversion from $M_b$ to $M_g$ for a rapidly rotating NS becomes essential\footnote{It is worth noticing that besides knowing the relations between $M_g$ and $M_b$, one still faces some problems such as extracting the ejecta mass information from EM observations or estimating the initial spin period in the rigidly rotation phase, especially when significant angular momentum loss (e.g. due to the strong viscous spin down, \citealt{radice18}) is considered.}. Finally, the maximum gravitational mass is significantly enhanced by rapid rotation. For some constraints on $M_{\rm TOV}$, the relation between the maximum NS mass supported through uniform rotation (marked as $M_{\rm max}$) and $M_{\rm TOV}$ is employed. 

In this work we consider rigidly rotating stellar equilibrium models (treated by numerical relativity methods) and aim to find relations between $M_g$ and $M_b$, and between $M_{\rm max}$ and $M_{\rm TOV}$, for a selected sample of EOSs, which are consistent with current observations (i.e., $M_{\rm TOV}$ is larger than $\sim2.01 M_{\odot}$, which is the the largest well-measured NS mass for PSR J0348+0432 \citep{antoniadis13}) and have a range of maximum masses. Here we only consider rigidly rotating non-magnetized NSs at zero temperature. NS structural quantities in the differential rotation phase has been discussed in previous works \cite[e.g.][]{studzinska16,gondek-Rosinska17,bauswein17,bozzola18,weih18}.

\section{NS Structure Equations and NS EOS}

We consider the equilibrium equations for a stationary, axially symmetric, rigid rotating NS, within a fully general relativistic framework. The spacetime metric can be written in the form
\begin{eqnarray}
ds^2 &=& -e^{2\nu}\, dt^2 + r^2 \sin^2\theta B^2 e^{-2\nu} 
(d\phi - \omega\, dt)^2 
\nonumber \\ & & \mbox{}
+ e^{2\alpha} (dr^2 + r^2\, d\theta^2),
\end{eqnarray}
where the potentials $\nu$, $B$, $\omega$, and $\alpha$ depend only on $r$ and $\theta$, and have the
following asymptotic decay \citep{butterworth76}
\begin{eqnarray}
\nu &=& -\frac{M}{r} + \frac{B_0M}{3r^3} + \nu_2 P_2(\cos\theta)+\mathcal{O}\left(\frac{1}{r^{4}}\right), \nonumber \\
B &=& 1 + \frac{B_0}{r^2} +\mathcal{O}\left(\frac{1}{r^{4}}\right) , \nonumber \\ 
\omega &=&\frac{2I\Omega}{r^3}+\mathcal{O}\left(\frac{1}{r^{4}}\right) ,
\end{eqnarray}
where $M$, $I$ and $\Omega$ are the NS mass, moment of inertia and angular frequency, respectively. $B_0$ and $\nu_2$ are real constants. 

We describe the interior of the NS as a perfect fluid, whose energy-momentum tensor becomes 
\begin{eqnarray}
T^{\mu\nu} = (\rho + P)u^{\mu}u^{\nu} + P g^{\mu\nu},
\end{eqnarray}
where $\rho$ and $P$ are the energy density and the pressure, and $u^{\mu}$ is the
$4$-velocity. Given a particular NS EOS, we use the public code \texttt{RNS} \citep{stergioulas95} to solve the field equations for the rotating NS. 

Our selection of realistic (tabulated) EOSs (as listed in Table 1) are SLy \citep{douchin01}, WFF1 \citep{wiringa88}, WFF2 \citep{wiringa88}, AP3 \citep{akmal97}, AP4 \citep{akmal97}, BSK21 \citep{goriely10}, DD2 \citep{typel10}, MPA1 \citep{muther87}, MS1 \citep{muller96}, MS1b \citep{muller96}, with $M_{\rm TOV}$ ranging from $2.05M_{\odot}$ to $2.78M_{\odot}$. 

\section{Relation between $M_g$ and $M_b$}

Given a particular EOS, the baryonic mass ($M_b$) and gravitational mass ($M_g$) for a rigid rotating NS are determined by the value of central energy density ($\rho_c$) and spin period ($P$). In order to figure out the relationship between $M_b$ and $M_g$, for each EOS, we calculated a series of $M_b$ and $M_g$ for different spin period. Here we present the results for $P$ equaling to \{1.3,1.4,1.5,1.6,1.7,1.8,1.9,2.0,3.0,4.0\}$P_k$, where $P_k$ is the keplerian period corresponding to $M_{\rm TOV}$. In the following we consider relations describing all EOSs and relations for individual EOSs, which may become useful if better constraints on the EOS become available.

\subsection{Non-rotating NS}
\begin{figure*}[t]
\begin{center}
\begin{tabular}{ll}
\resizebox{80mm}{!}{\includegraphics[]{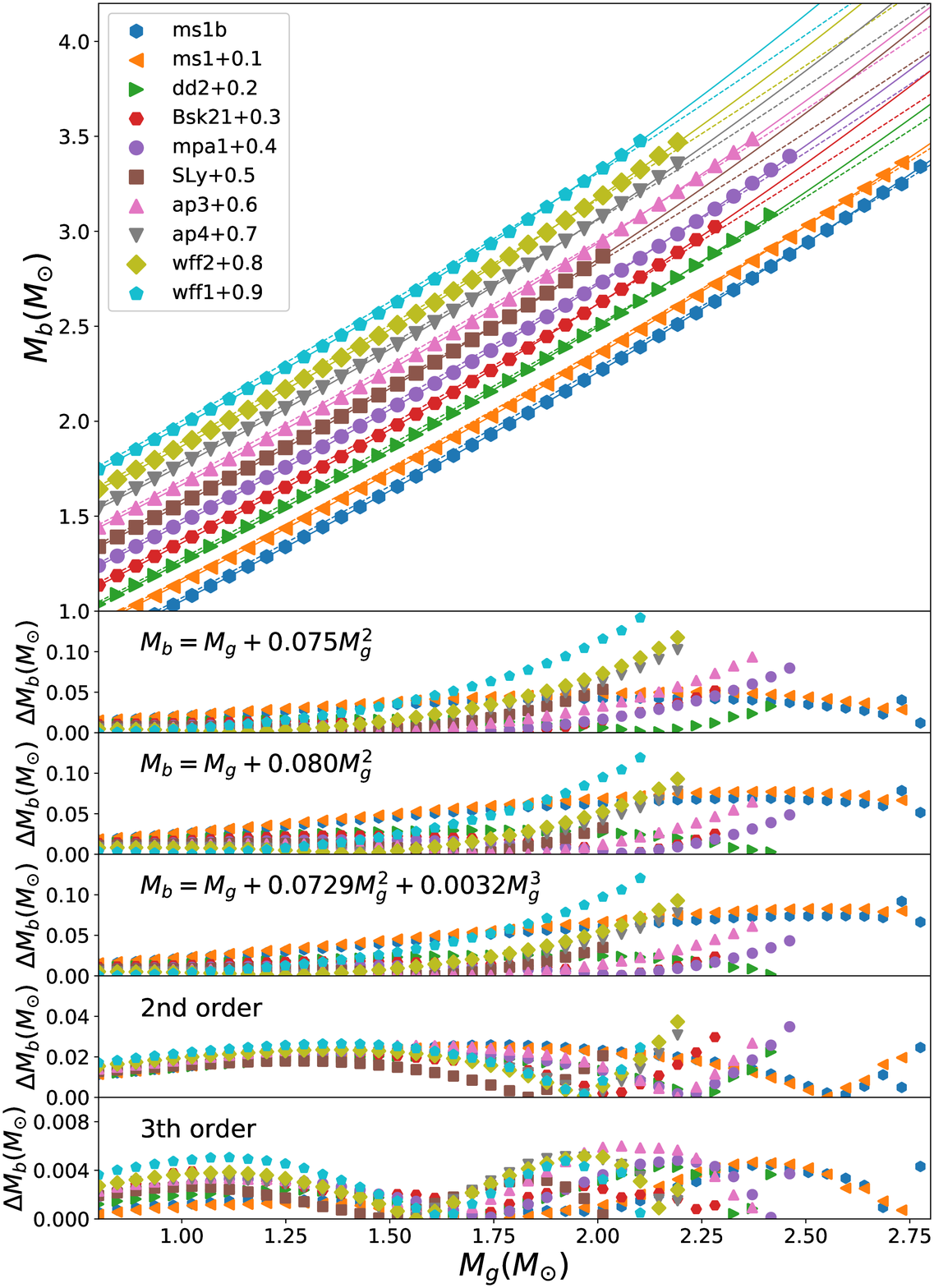}} &
\resizebox{80mm}{!}{\includegraphics[]{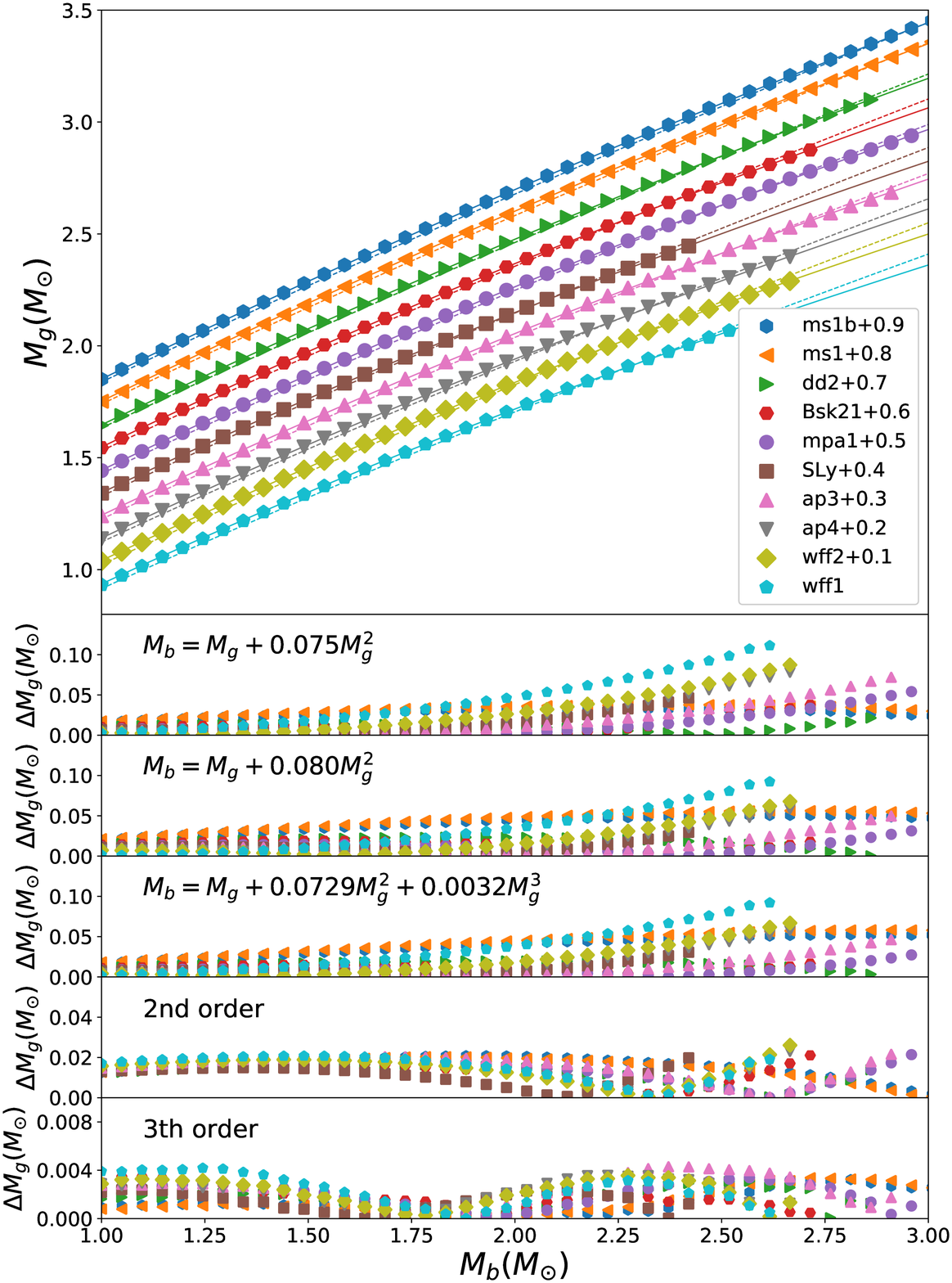}} \\
\end{tabular}
\caption{The relation between the baryonic mass ($M_b$) and gravitational mass ($M_g$) for a non-rotating NS and the residual error for different transformation formulae. Different colors denote different EOSs. The solid and dashed lines represent the best fitting results with the quadratic and cubic formulae, respectively, with the best-fit parameters adopted for each EOS.}
\label{fig:non-rot}
\end{center}
\end{figure*}

For a non-rotating NS, transformation of the baryonic NS mass to gravitational mass is commonly approximated using the quadratic formula
\begin{eqnarray}
M_b=M_g+A\times M_{g}^2,
\end{eqnarray}
where $A$ is usually adopted as a constant number 0.075 \citep{timmes96}. Throughout the paper, $M_b$ and $M_g$ are in units of $M_\odot$. Here we treat $A$ as a free parameter in order to find out its best fit value for each adopted EOS (results are collected in Table 1). If each EOS adopting its own best fit $A$ value, the residual error of the transformation between $M_g$ and $M_b$ is in order of a few $10^{-2}M_\odot$, see Figure \ref{fig:non-rot} for details. Putting ($M_g$, $M_b$) results for all adopted EOS together, the overall best fit $A$ value is 0.080. In this case, the residual error of the transformation from $M_g$ to $M_b$ is as large as 0.12$M_\odot$ and the residual error from $M_b$ to $M_g$ is as large as 0.09$M_\odot$. By comparison, when $A$ is adopted as 0.075, the residual error of the transformation from $M_g$ to $M_b$ is as large as 0.14$M_\odot$ and the residual error from $M_b$ to $M_g$ is as large as 0.11$M_\odot$. 

\begin{table*}\
\begin{center}
\caption{Characteristic parameters for various EOSs.}
\begin{tabular}{ccccccccccc}
  \hline
 \hline

        &$M_{\rm TOV}$ &$R_{1.4}$ &A\tablenotemark{*}    &$A_1$\tablenotemark{*}   &$A_2$\tablenotemark{*}    & $P_k$       &$\alpha$                              &$\beta$ & a &b\\
        &$\left(M_{\odot}\right)$&$({\rm km})$ &~&~&~        & $({\rm ms})$&$\left(10^{-10}{\rm s}^{-\beta}\right)$&~&~&~\\
\hline
SLy     & 2.05          & 11.69   & 0.083 & 0.041 & 0.023    & 0.55 & 2.311 & -2.734  & 0.086 & 0.209\\  
\hline                                                                                                   
WWF1    & 2.14          & 10.40   & 0.102 & 0.046 & 0.029    & 0.47 & 3.111 & -2.729  & 0.106 & 0.279\\  
\hline                                                                                                   
WWF2    & 2.20          & 11.1    & 0.092 & 0.043 & 0.026    & 0.50 & 2.170 & -2.694  & 0.096 & 0.270\\  
\hline                                                                                                   
AP4     & 2.22          & 11.36   & 0.090 & 0.045 & 0.023    & 0.51 & 2.095 & -2.721  & 0.094 & 0.251\\  
\hline                                                                                                   
BSK21   & 2.28          & 12.55   & 0.079 & 0.039 & 0.020    & 0.60 & 1.958 & -2.799  & 0.083 & 0.257\\  
\hline                                                                                                   
AP3     & 2.39          & 12.01   & 0.087 & 0.046 & 0.019    & 0.55 & 1.883 & -2.780  & 0.091 & 0.288\\  
\hline                                                                                                   
DD2     & 2.42          & 13.12   & 0.077 & 0.046 & 0.014    & 0.65 & 2.269 &--2.811  & 0.079 & 0.259\\  
\hline                                                                                                   
MPA1    & 2.48          & 12.41   & 0.082 & 0.046 & 0.017    & 0.59 & 2.584 & -2.693  & 0.088 & 0.284\\  
\hline                                                                                                   
Ms1     & 2.77          & 14.70   & 0.069 & 0.042 & 0.010    & 0.72 & 5.879 & -2.716  & 0.072 & 0.263\\  
\hline                                                                                                   
Ms1b    & 2.78          & 14.46   & 0.070 & 0.043 & 0.011    & 0.71 & 4.088 & -2.770  & 0.074 & 0.283\\  

\hline
\hline
\end{tabular}
\tablenotetext{*}{Best fit vales for non-spinning NS cases.}
\end{center}
\end{table*}

In order to further reduce the residual error of the transformation, we test a new cubic formula
\begin{eqnarray}
M_b=M_g+A_1\times M_{g}^2+A_2\times M_{g}^3.
\end{eqnarray}
We determine the best fit values of $A_1$ and $A_2$ for each adopted EOS (collected in table 1). If each EOS adopting its own best-fit values, the residual error of the transformation between $M_g$ and $M_b$ is in the order of a few $10^{-3} M_\odot$, which is almost one order of magnitude better than the quadratic cases. See Figure \ref{fig:non-rot} for details. Putting all EOSs together, the overall best fit $A_1$ and $A_2$ values are 0.0729 and 0.0032. In this case, the residual error of the transformation from $M_g$ to $M_b$ is as large as 0.12$M_\odot$ and the residual error from $M_b$ to $M_g$ is as large as 0.09$M_\odot$, which are in the same order as the quadratic case. 

In the literature, some universal relations between the binding energy $M_b-M_g$ and the neutron star's compactness have been proposed \citep{lattimer01,coughlin17}, which are also applicable for transformation between $M_b$ and $M_g$. Here we find that for the quadratic transformation, one has $A\times R_{1.4}\approx 1$ for all adopted EOS, where $A$ is the best fit value and $R_{1.4}$ is the NS radius (in unit of km) for 1.4$M_\odot$. We thus propose a new universal relation
\begin{eqnarray}
M_b=M_g+R^{-1}_{1.4}\times M_{g}^2,
\end{eqnarray}
with $1.8\%$ relative error. We note that in our proposed relation, only $R_{1.4}$ is invoked in the relation, instead of using the compactness number proposed in previous works \cite[e.g.][]{lattimer01,coughlin17}. The accuracy is similar to the previous ones in terms of the residual error of the transformation, but our proposed relation is more practical since it does not involve the calculations of the radii corresponding to different masses. In case that the binding energy $M_b-M_g$ is measured independently, 
such a universal relation may be helpful to constrain $R_{1.4}$, which helps to distinguish the NS EOSs \citep{lattimer01}.

\begin{figure}[t!] 
\vspace{0.3cm}
{\centering
{\includegraphics[width=3.2in]{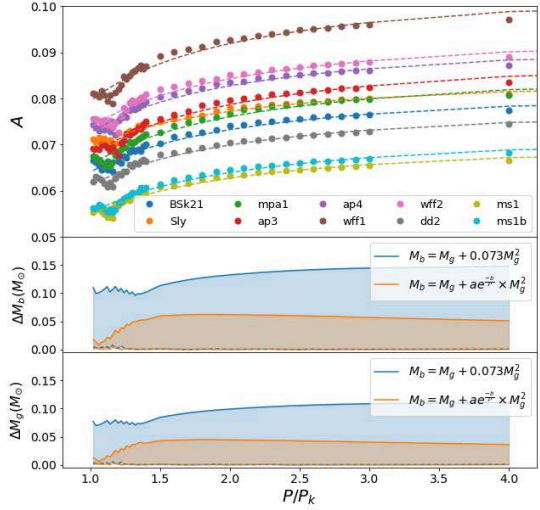}}
\par}
\label{fig:Ap}
\caption{Relation between the quadratic transformation coefficient A and the spin period of the NS. The dashed lines represent the best fitting results with formula $A \approx ae^{-{b \over {\cal P}}}$, with the fitting uncertainty showing in the lower panels.}
\end{figure}

\begin{figure}[t]
\begin{center}
\begin{tabular}{ll}
\includegraphics[width=3.2in]{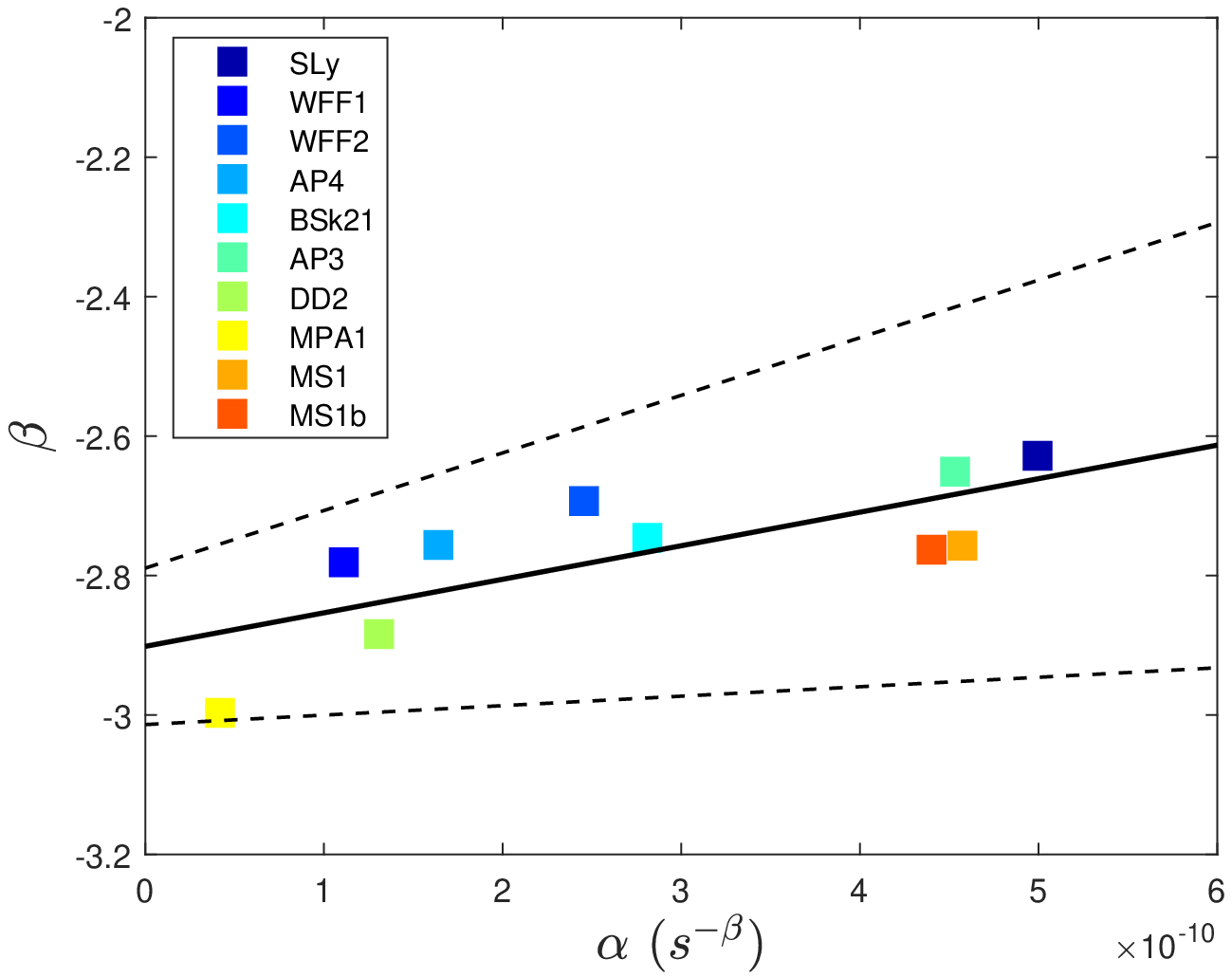}  \\
\includegraphics[width=3.2in]{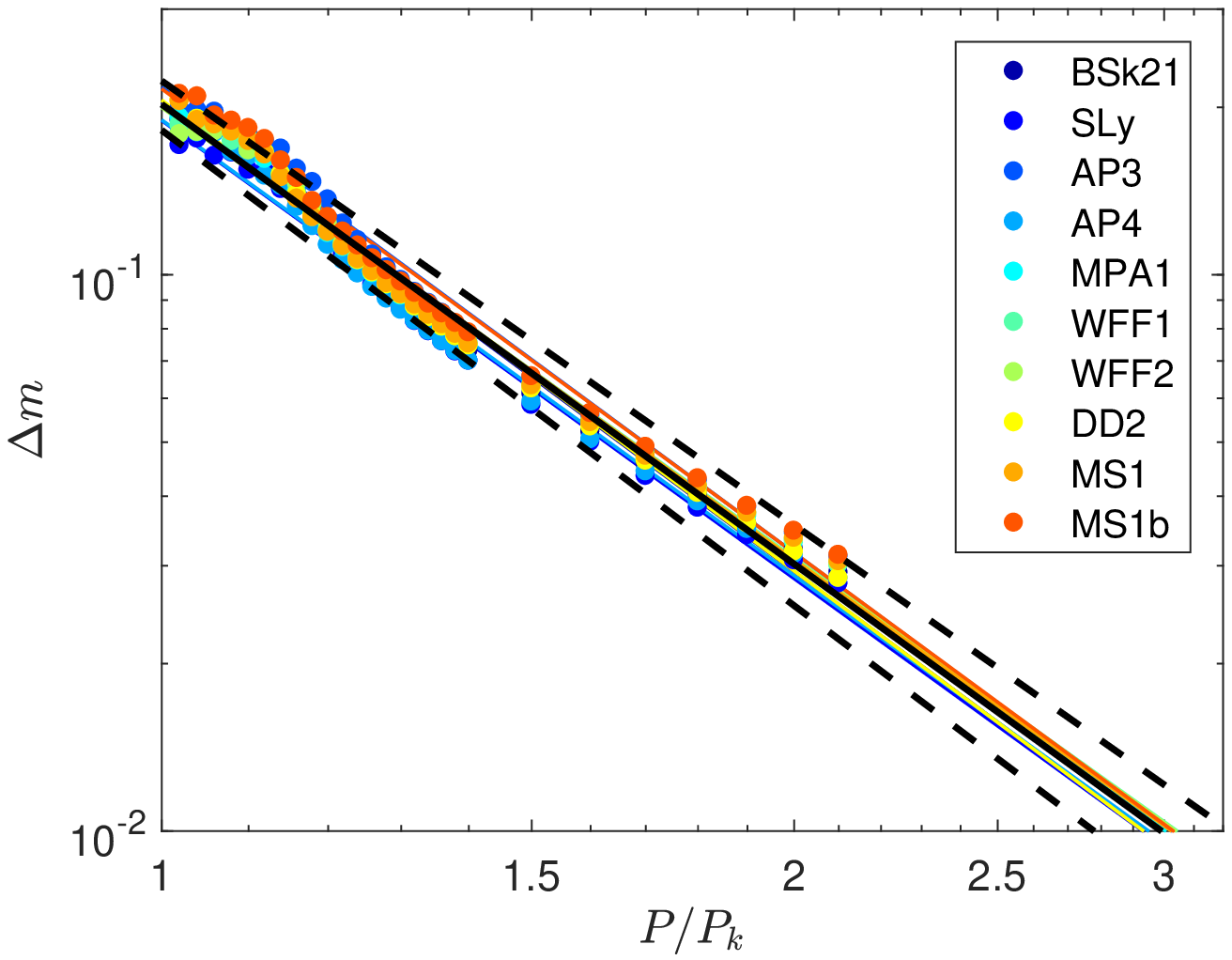} \\
\end{tabular}
\caption{Upper panel: The correlation between $\alpha$ and $\beta$ for our selected EOSs. The solid and dashed lines present the regression line and its 3-$\sigma$ region. The correlation coefficient between $\alpha$ and $\beta$ is 0.75. Lower panel: The correlation between the normalized mass gain due to spin $\Delta m\equiv(M_{\rm max}-M_{\rm TOV})/M_{\rm TOV}$ and the normalized spin period ${\cal P}\equiv P/P_{K}$. The thin colored solid lines represent the results when specific $\alpha$ and $\beta$ values are adopted for each EOS. The thick black solid and dashed lines represent the universal relation proposed by equation \ref{fp} and its 2-$\sigma$ region.} 
\label{fig:fp}
\end{center}
\end{figure}

\subsection{Rotating NS}

For individual EOSs, we also fit their $M_b$ and $M_g$ relations with both the quadratic and cubic formulae for the situations with different rotation periods. We find following conclusions which could be applied to all adopted EOSs: 1) The relation between $M_b$ and $M_g$ is different for different spin periods. 2) If one adopts the best-fit parameters for each spin period, the cubic formula is almost one order of magnitude better than the quadratic formula in terms of the residual error of the transformation. For instance, the residual error of the transformation from $M_g$ to $M_b$ is up to 0.06$M_\odot$ for the quadratic case and is up to 0.01$M_\odot$ for the cubic case. 3) Putting together the ($M_g$, $M_b$) results for different spin periods, with the overall best-fit values, the cubic formula is no better than the quadratic formula. For instance, the residual error of the transformation from $M_g$ to $M_b$ is up to 0.068$M_\odot$ for the quadratic case and is up to 0.062$M_\odot$ for the cubic case. For a comparison, when $A$ is fixed to 0.075, the residual error of the transformation from $M_g$ to $M_b$ is as large as 0.11$M_\odot$. The best-fit values for the quadratic and cubic formulae are collected in Table 2. 

Considering ($M_g$, $M_b$) results for different EOSs and different spin periods together, the overall best-fit value is $A=0.073$ for the quadratic formula and $A_1=0.0777$ and $A_2=0.0018$ for the cubic formula. In this case, the residual error of the transformation from $M_g$ to $M_b$ is up to 0.18$M_\odot$ for the quadratic case and is up to 0.16$M_\odot$ for the cubic case. It is interesting to note that the overall best fit $A$ value is very close to 0.075 as proposed in \cite{timmes96}

For a given EOS, the quadratic transformation parameter $A$ is a rough function of the spin period (see Figure 2),
\begin{eqnarray}
A \approx ae^{-{b \over {\cal P}}},
\end{eqnarray}
where ${\cal P}\equiv P/P_{K}$ is the characterized NS spin period normalized to the Keplerian period. The best-fit values for $a$ and $b$ are collected in Table 1. Interestingly, we find that $a\times R_{1.4}\approx 1$ and $b\approx1/4$ for all the adopted EOSs. We thus have a new universal relation for a rotating NS,
\begin{eqnarray}
M_b=M_g+R^{-1}_{1.4}e^{-{1 \over 4 {\cal P}}}\times M_{g}^2,
\end{eqnarray}
with a $3.3\%$ relative error. Two EOS related parameters $R_{1.4}$ and $P_k$ are invoked in this relation, which may be constrained if $M_b$ and the binding energy $M_b-M_g$ for a NS could be independently measured.

\section{$M_{\rm max}$ and $M_{\rm TOV}$ relation}

It has been proposed to parameterize $M_{\rm max}$ as a function of the spin period of the central star \citep{lasky14},
\begin{eqnarray}
M_{\rm max} = M_{\rm TOV}(1+\alpha P^{\beta}),
\label{Mt1}
\end{eqnarray}
where $P$ is the spin period of the NS in units of second. Using the {\tt RNS} code, we calculate the numerical values for $\alpha$ and $\beta$ for our adopted EOSs. The results are presented in Table 1. 

We find that $\alpha$ and $\beta$ are not independent of each other (see Figure 3 for details). If we define 
\begin{eqnarray}
\Delta m \equiv(M_{\rm max}-M_{\rm TOV})/M_{\rm TOV}=\alpha P^{\beta},
\label{deltaM}
\end{eqnarray}
as the normalized mass gain due to spin, as shown in figure 3, a very tight and general correlation between $\Delta m$ and ${\cal P}$ could be found as
\begin{eqnarray}
{\rm log_{10}}\Delta m = (-2.74\pm0.05){\rm log_{10}}{\cal P}+{\rm log_{10}}(0.20\pm 0.01),
\label{fp}
\end{eqnarray}
which is essentially independent of the EOS models. It is worth noticing that when ${\cal P}=1$, i.e., when NS spin period equals to the Keplerian period, one has $\Delta m \simeq 0.2$, or $M_{\rm max}\simeq1.2M_{\rm TOV}$, which is well consistent with the previous numerical simulation results \citep{cook94,lasota96,breu16}.

\begin{table*}
\begin{center}
\caption{Best fitting results for different EoSs with different spin periods.}
\begin{tabular}{c|ccc|ccc|ccc|ccc|ccc}
  \hline
\hline
~          &~      & SLy    &~           &~        & WFF1   &~        &~        & WFF2   &~        &~        &AP4     &~         &~        &BSk21   &~       \\         
 \hline
~          &$A$& $A_{1}$& $A_{2}$    &$A$& $A_{1}$& $A_{2}$ &$A$& $A_{1}$& $A_{2}$ &$A$& $A_{1}$& $A_{2}$  &$A$& $A_{1}$& $A_{2}$\\ 
\hline                                                                                                                                                          
$P=1.3P_k$ & 0.072 & -0.318 & 0.177      & 0.085 & -0.172 & 0.112     & 0.079 & -0.327 & 0.171     & 0.077 & -0.214 & 0.123      & 0.068 & -0.179 & 0.101 \\    
\hline                                                                                                                                                          
$P=1.4P_k$ & 0.073 & -0.112 & 0.087      & 0.084 & -0.058 & 0.066     & 0.079 & -0.111 & 0.083     & 0.077 & -0.075 & 0.067      & 0.068 & -0.058 & 0.054 \\    
\hline                                                                                                                                                          
$P=1.5P_k$ & 0.072 & -0.046 & 0.058      & 0.085 & -0.020 & 0.050     & 0.080 & -0.049 & 0.058     & 0.078 & -0.029 & 0.049      & 0.067 & -0.017 & 0.038 \\    
\hline                                                                                                                                                          
$P=1.6P_k$ & 0.073 & -0.016 & 0.045      & 0.086 & 0.002 & 0.042      & 0.079 & -0.017 & 0.045     & 0.077 & -0.004 & 0.039      & 0.068 & 0.001 & 0.032 \\     
\hline                                                                                                                                                          
$P=1.7P_k$ & 0.073 & 0.000 & 0.038       & 0.086 & 0.015 & 0.037      & 0.079 & 0.000 & 0.039      & 0.078 & 0.009 & 0.034       & 0.068 & 0.012 & 0.028 \\     
\hline                                                                                                                                                          
$P=1.8P_k$ & 0.073 & 0.011 & 0.034       & 0.086 & 0.024 & 0.033      & 0.079 & 0.012 & 0.034      & 0.078 & 0.018 & 0.031       & 0.069 & 0.019 & 0.025 \\     
\hline                                                                                                                                                          
$P=1.9P_k$ & 0.073 & 0.018 & 0.031       & 0.088 & 0.028 & 0.032      & 0.080 & 0.019 & 0.031      & 0.079 & 0.023 & 0.029       & 0.069 & 0.024 & 0.023 \\     
\hline                                                                                                                                                          
$P=2.0P_k$ & 0.074 & 0.022 & 0.029       & 0.088 & 0.033 & 0.030      & 0.081 & 0.024 & 0.030      & 0.079 & 0.028 & 0.026       & 0.069 & 0.028 & 0.022 \\     
\hline                                                                                                                                                          
$P=3.0P_k$ & 0.076 & 0.038 & 0.023       & 0.091 & 0.047 & 0.026      & 0.083 & 0.041 & 0.024      & 0.082 & 0.043 & 0.022       & 0.072 & 0.037 & 0.019 \\     
\hline                                                                                                                                                          
$P=4.0P_k$ & 0.077 & 0.041 & 0.022       & 0.093 & 0.049 & 0.026      & 0.085 & 0.044 & 0.024      & 0.083 & 0.046 & 0.022       & 0.073 & 0.039 & 0.019 \\     
\hline
\hline
~          &~      & AP3    &~           &~        & DD2    &~        &~        & MPA1   &~        &~        &MS1     &~         &~        &MS1b    &~       \\
 \hline                                                                                                                                                        
~          &$A$& $A_{1}$& $A_{2}$    &$A$& $A_{1}$& $A_{2}$ &$A$& $A_{1}$& $A_{2}$ &$A$& $A_{1}$& $A_{2}$  &$A$& $A_{1}$& $A_{2}$\\
\hline                                                                                                                                                         
$P=1.3P_k$ & 0.072 & -0.292 & 0.140      & 0.064   & -0.100 & 0.064   & 0.069   & -0.112 & 0.070   & 0.058   & -0.095 & 0.052    & 0.059 & -0.104   & 0.055 \\ 
\hline                                                                                                                                                         
$P=1.4P_k$ & 0.073 & -0.095 & 0.068      & 0.064   & -0.028 & 0.038   & 0.070   & -0.039 & 0.043   & 0.058   & -0.030 & 0.031    & 0.059 & -0.031   & 0.032 \\ 
\hline                                                                                                                                                         
$P=1.5P_k$ & 0.073 & -0.033 & 0.045      & 0.065   & -0.002 & 0.029   & 0.070   & -0.005 & 0.032   & 0.058   & -0.006 & 0.024    & 0.059 & -0.004   & 0.023 \\ 
\hline                                                                                                                                                         
$P=1.6P_k$ & 0.073 & -0.006 & 0.035      & 0.065   & 0.013  & 0.023   & 0.070   & 0.010  & 0.027   & 0.058   & 0.007 & 0.020     & 0.059 & 0.009    & 0.019 \\ 
\hline                                                                                                                                                         
$P=1.7P_k$ & 0.074 & 0.008 & 0.030       & 0.066   & 0.022  & 0.020   & 0.071   & 0.019  & 0.023   & 0.059   & 0.015 & 0.017     & 0.060 & 0.017    & 0.017 \\ 
\hline                                                                                                                                                         
$P=1.8P_k$ & 0.074 & 0.017 & 0.027       & 0.066   & 0.027  & 0.019   & 0.071   & 0.026  & 0.021   & 0.059   & 0.020 & 0.016     & 0.060 & 0.023    & 0.015 \\ 
\hline                                                                                                                                                         
$P=1.9P_k$ & 0.074 & 0.024 & 0.024       & 0.066   & 0.031  & 0.017   & 0.072   & 0.031  & 0.020   & 0.060   & 0.024 & 0.015     & 0.061 & 0.027    & 0.014 \\ 
\hline                                                                                                                                                         
$P=2.0P_k$ & 0.075 & 0.029 & 0.023       & 0.067   & 0.034  & 0.017   & 0.073   & 0.033  & 0.019   & 0.060   & 0.027 & 0.015     & 0.061 & 0.030    & 0.013 \\ 
\hline                                                                                                                                                         
$P=3.0P_k$ & 0.078 & 0.043 & 0.018       & 0.070   & 0.043  & 0.014   & 0.076   & 0.044  & 0.016   & 0.062   & 0.037 & 0.012     & 0.064 & 0.039    & 0.011 \\ 
\hline                                                                                                                                                         
$P=4.0P_k$ & 0.080 & 0.046 & 0.018       & 0.071   & 0.045  & 0.014   & 0.077   & 0.046  & 0.016   & 0.064   & 0.039 & 0.011     & 0.065 & 0.041    & 0.011 \\ 
\hline
\hline
\end{tabular}
\end{center}
\label{table:fit}
\end{table*}

\section{Conclusion and Discussion}
In this work, we solved the field equations for the rotating NS with a selected sample of EOSs. For each EOS, we calculated a series of $M_b$ and $M_g$ for different spin periods. Our results could be summarized as follows:

\begin{itemize}
\item For non-rotating NSs, if one intends to apply an EOS-independent universal quadratic or cubic transformation formula to all the EOSs, one has the best-fit formula $M_b=M_g+0.080 M_g^2$ and $M_b = M_g + 0.0729 M_g^2 + 0.0032 M_g^3$. The residual error of the transformation from $M_g$ to $M_b$ is as large as 0.12$M_\odot$, and the residual error from $M_b$ to $M_g$ is as large as 0.09$M_\odot$. There is no advantage for the higher order formula. However, for individual EOSs, if one adopts its own best-fit values for the coefficients, the cubic formula is much better than the quadratic formula, with the residual error of the transformation less than 0.01$M_\odot$.
\item For a rotating NS, the relation between $M_b$ and $M_g$ is different for different spin periods. If one intends to apply an EOS-independent universal quadratic or cubic transformation formula to all the EOSs for all spin periods, one has $M_b=M_g+0.073 M_g^2$ and $M_b = M_g + 0.0777 M_g^2 + 0.0018 M_g^3$. The residual error of the transformation is up to 0.18$M_\odot$. There is still no advantage for the higher order formula. Given an EOS, if one intends to apply an EOS-independent universal quadratic or cubic transformation formula to all spin periods, the residual error of the transformation is up to 0.068$M_\odot$. Again no advantage for the higher order formula. However, if one adopts the best-fit parameters for each spin period, the cubic transformation residual error is less than 0.01$M_\odot$, almost one order of magnitude better than the quadratic formula.
\item For quadratic transformation, we find two EOS insensitive relations: $M_b=M_g+R^{-1}_{1.4}\times M_{g}^2$ for a non-rotating NS and $M_b=M_g+R^{-1}_{1.4}e^{-{1 \over 4 {\cal P}}}\times M_{g}^2$ for rotating NSs with spin period $P={\cal P}\times P_k$. In principle, the former relation for non-rotating NS (which may be also applied to slowly spinning NS) could be used to make constraint on NS radius $R_{1.4}$, thus the NS equation of state, once the $M_b$ and $M_g$ for a NS could be independently measured \citep{lattimer01}. It becomes less straightforward for the relation including spin, since two EOS related parameters $R_{1.4}$ and $P_k$ are invoked and the NS spin period may be difficult to determine.
\item With our calculations, we also find a very tight and general correlation between the normalized mass gain due to spin $\Delta m\equiv(M_{\rm max}-M_{\rm TOV})/M_{\rm TOV}$ and normalized spin period ${\cal P}$ as ${\rm log_{10}}\Delta m = (-2.74\pm0.05){\rm log_{10}}{\cal P}+{\rm log_{10}}(0.20\pm 0.01)$, which is independent of EOS models. Note that this universal relation is only valid for rigidly rotating NSs. 
\end{itemize}

It has been proposed that binary neutron star merger events could give tight constraints on the NS maximum mass, as long as we could calculate the mass of the merger remnant and justify whether the remnant is a BH or a long-lasting NS. For instance, it has been claimed that the multi-messenger observations of GW170817 could provide an upper bound on $M_{\rm TOV}$ \citep{margalit17,rezzolla18,ruiz18}, e.g. $M_{\rm TOV}\lesssim2.16~M_{\odot}$ in \citep{margalit17}. However, this result sensitively depends on the assumption that the merger remnant of GW170817 is a not-too-long-lived hypermassive NS. This is still subject to debate, see supporting arguments by \cite{margalit17,rezzolla18,ruiz18} and counter opinion by \cite{yu18,li18,ai18,piro19}. This is due to the lack of GW detection in the post-merger phase and the insufficient capability to distinguish the merger product using electromagnetic (EM) observations only. If the low-significance flare-like feature at 155 days in the X-ray afterglow of GW170817 as claimed in \cite{piro19} is true, or if the argument by \cite{li18} that an additional energy injection from the merger product is required to interpret the blue component of AT2017gfo (GW170817 optical counterpart) is valid, the central remnant of GW170817 has to be a long-lived NS. The constraints on $M_{\rm TOV}$ would be reversed to a lower bound. Note that before GW170817 was detected, it was proposed that the statistical observational properties of Swift SGRBs favors NS EOS with $M_{\rm TOV}$ greater than $2.2~M_{\odot}$ \citep{gao16,li16}, if the cosmological NS mass distribution follows that observed in the BNS systems in our Galaxy.

In the case that the merger remnant of GW170817 was a long-lasting, rigidly rotating NS, the remnant gravitational mass could be estimated as follows: the total gravitational mass of the binary system is determined as $m_{g,1}+m_{g,2}=2.74M_{\odot}$ based on the inspiral phase GW signal \citep{abbott17}. The mass ratio of the binary is bound to 0.7-1 under the low dimensionless NS spin prior. Without any prior for the NS EOS, the total baryonic mass could be estimated as\footnote{The total baryonic mass is highly insensitive to the binary mass ratio. For instance, the total baryonic mass for GW170817 is $3.04\pm0.10M_\odot$ for $m_{g,1}/m_{g,2}=1$ and $3.05\pm0.10M_\odot$ for $m_{g,1}/m_{g,2}=0.7$.} $3.04M_\odot$ by using the EOS-independent universal quadratic formula for non-rotating NSs (i.e. $M_b=M_g+0.08\times M_{g}^2$). The uncertainty for the transformation is up to 0.10$M_\odot$. With the EM counterpart observations, the ejecta mass for GW170817 is estimated as $\sim0.06M_\odot$, with uncertainty of the order of several $10^{-2}M_\odot$ \cite[][and reference therein]{metzger17}. In this case, the baryonic mass for the merger remnant could be estimated as $2.98M_{\odot}$, with conservative uncertainty up to 0.2$M_\odot$. If the remnant is a rapidly rotating NS, although it is difficult to determine its inital spin period, we can transform its baryonic mass to the gravitational mass as $2.52M_{\odot}$, by using the EOS-independent universal quadratic formula for arbitrary rotating NSs (i.e. $M_b=M_g+0.073\times M_{g}^2$), with the transformation uncertainty being up to 0.10$M_\odot$. Putting together all the uncertainties in the transformation\footnote{We also consider that some errors will cancel out in the process of interconversion between $M_g$ and $M_b$.} and the ejecta mass, the overall uncertainty for $M_g$ of the merger remnant could be up to $\sim~0.15~M_\odot$.  Recently, \cite{radice18} proposed that given an EOS, the initial spin period of the merger remnant could be estimated based on the value of its baryonic mass. In this case, if we have some strong prior for the NS EOS, the transformation uncertainty between $M_b$ and $M_g$ could be reduced to the order of 0.001$M_\odot$, if one applies the cubic transformation formula with the best fit values for a specific EOS with a specific initial spin period. The overall uncertainty for this case is mainly contributed by the ejecta mass uncertainty, which might be further reduced in the future when a larger sample of EM counterparts for binary NS mergers are observed.

\acknowledgments

This work is supported by the National Natural Science Foundation of China under Grant No. 11722324, 11603003, 11633001, 11690024 and 11873040, the Strategic Priority Research Program of the Chinese Academy of Sciences, Grant No. XDB23040100 and the Fundamental Research Funds for the Central Universities. A.B. acknowledges support by the European Research Council (ERC) under the European Union's Horizon 2020 research and innovation programme under grant agreement No. 759253 and by the Sonderforschungsbereich SFB 881 ``The Milky WaySystem" (subproject A10) of the German Research Foundation (DFG).

{}

\end{document}